\documentclass[pra,aps,showpacs,groupedaddress,superscriptaddress,twocolumn]{revtex4-1}
\usepackage[utf8x]{inputenc}
\usepackage{color}
\usepackage{bbm} 

\usepackage{amsfonts,amsmath,amssymb,stmaryrd}

\usepackage{graphicx}
\usepackage{subfigure}  
\usepackage{bbm} 
\usepackage[pdftex]{epsfig}
\usepackage{verbatim}
\usepackage{siunitx}
\begin{document}

\title{Direct laser written polymer waveguides with out of plane couplers for optical chips}

\author{Alexander Landowski}
\affiliation{Department of Physics and State Research Center OPTIMAS, University of Kaiserslautern, Erwin-Schroedinger-Str. 46, 67663 Kaiserslautern, Germany}
\affiliation{Graduate School Materials Science in Mainz, Erwin-Schroedinger-Str. 46, 67663 Kaiserslautern, Germany}
\author{Dominik Zepp}
\affiliation{Department of Physics and State Research Center OPTIMAS, University of Kaiserslautern, Erwin-Schroedinger-Str. 46, 67663 Kaiserslautern, Germany}
\author{Sebastian Wingerter}
\affiliation{Department of Physics and State Research Center OPTIMAS, University of Kaiserslautern, Erwin-Schroedinger-Str. 46, 67663 Kaiserslautern, Germany}
\author{Georg von Freymann}
\affiliation{Department of Physics and State Research Center OPTIMAS, University of Kaiserslautern, Erwin-Schroedinger-Str. 46, 67663 Kaiserslautern, Germany}
\affiliation{Fraunhofer Institute for Industrial Mathematics ITWM, Fraunhofer-Platz 1, 67663 Kaiserslautern}
\author{Artur Widera}
\affiliation{Department of Physics and State Research Center OPTIMAS, University of Kaiserslautern, Erwin-Schroedinger-Str. 46, 67663 Kaiserslautern, Germany}
\affiliation{Graduate School Materials Science in Mainz, Erwin-Schroedinger-Str. 46, 67663 Kaiserslautern, Germany}



\begin{abstract}
Optical technologies call for waveguide networks featuring high integration densities, low losses, and simple operation.
Here, we present polymer waveguides fabricated from a negative tone photoresist via two-photon-lithography in direct laser writing, and show a detailed parameter study of their performance.
Specifically, we produce waveguides featuring bend radii down to $\SI{40}{\micro\metre}$, insertion losses of the order of $10\,$dB, and loss coefficients smaller than \SI{0.81}{\decibel \per \milli\metre}, facilitating high integration densities in writing fields of $\SI{300}{\micro\metre}\times\SI{300}{\micro\metre}$.
A novel three-dimensional coupler design allows for coupling control as well as direct observation of outputs in a single field of view through a microscope objective.
Finally, we present beam-splitting devices to construct larger optical networks, and we show that the waveguide material is compatible with the integration of quantum emitters.
\end{abstract}
\maketitle

\section{Introduction}
Applications of micro-optical systems for sensing or interferometry as well as the emerging field of quantum technology call for optical platforms yielding tight control over light propagation at ever smaller feature sizes. 
A successful standard route to micro-optical applications are nanofabricated waveguides on chips \cite{Bogdanov.2017}, allowing, for example, the observation of quantum walks of correlated photons in two-dimensional waveguides \cite{Poulios.2014} or first approaches to optical quantum information processing \cite{Politi.2009}.
In order to extend optical microsystems to three dimensions, the method of direct laser writing (DLW) offers structuring of waveguides in glass and recently also diamond \cite{Hadden.20170120} for, e.\,g., boson sampling \cite{Crespi.2013} or simulating environment-assisted quantum transport \cite{Biggerstaff.2016}. 
Due to a small change in the index of refraction between waveguide core and surrounding medium, such waveguides can guide light and allow for evanescent wave coupling \cite{Szameit.2007}, but at the same time are restricted to feature sizes large compared to the wavelength. 
Beyond small changes within an optical medium, DLW allows for writing three-dimensional polymer structures that have been used, for example, to connect two photonic chips via laser-written interconnects \cite{Lindenmann.2015}.

Combining such waveguides with quantum emitters is an important step toward quantum optical applications.
Nitrogen-vacancy-centers (NVC) in diamond, for example, have been shown to emit single photons at room temperature \cite{Gruber.1997}, to yield coherent access to a single electron's spin \cite{Jelezko.2004}, and to violate, in a loophole-free setup, Bell's inequality \cite{Hensen.2015}.
Quantum emitters have been integrated into micro-optical networks by structuring waveguides in the same substrate also housing quantum emitters, such as diamond \cite{RiedrichMoller.2014,Mouradian.2015,Momenzadeh.2015} or semiconductor materials \cite{Javadi.2015}; or, alternatively, by immersing pre-characterized quantum emitters embedded in nanoparticles into waveguides, highly facilitated by the availability of pick-and-place schemes \cite{Schell.2011,Liebermeister.2014,LarsLiebermeister.2016,Shi.2016,Schell.2014,Sartison.2017,Faez.2014}.
\begin{figure*}[hbt]
	\centering
	\includegraphics{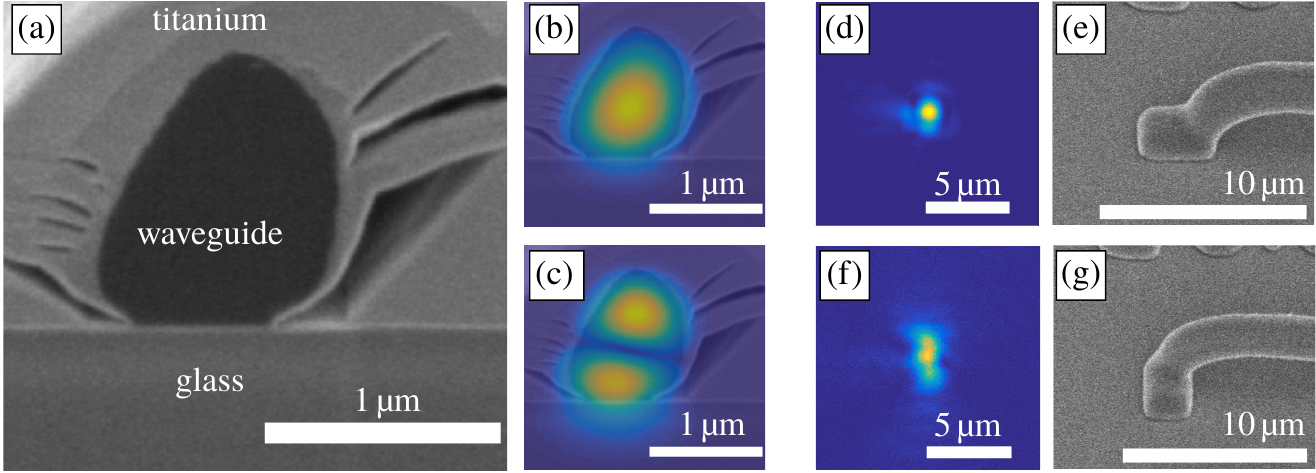}
	\caption{(a) Cross section of a waveguide, obtained by cutting the polymer waveguide on glass substrate with a focused ion beam and subsequent SEM under \SI{52}{\degree}, 
		in order to increase SEM contrast, the waveguides have been covered with titanium with a large growth rate, leading to gaps not reflecting the waveguide surface.
		(b),(c) Simulated intensity distribution of the two guided TE modes overlaid with the corresponding waveguide cross section, (d),(f) measured intensity distribution emitted from the outcoupling structure shown in (e),(g), respectively. The laser incoupling position was chosen to yield maximum transmission.
	} \label{fib}
\end{figure*}

Here, we present the fabrication and detailed parameter studies of three-dimensional optical waveguides with high refractive index contrast showing tight mode-guiding in a low-loss polymer on glass substrates.
Our waveguides go beyond existing structures in several respect. 
First, DLW allows combining extended planar sections laid onto the substrate with three dimensional structures. 
Consequently our waveguides allow combining large scale, stable networks with, e.\,g., out-of-plane coupling. 
Second, we exploit the latter to realize a novel coupler design facilitating simultaneous addressing and imaging of waveguide in- and outputs through the glass substrate using a single microscope objective, 
while all other sides of the substrate are still available for further manipulation.	
This enables, for example, controlled addressing and simultaneous direct observation of complex quantum optical experiments in one field of view of a microscope objective.
Third, due to the high refractive index contrast between waveguide and air, the mode field is tightly confined by the waveguide.
Finally, we show that the waveguides are in principle capable of integrating individual quantum emitters, such as NVC in nano-diamonds.

We engineer structure sizes small compared to the waveguide length, such as bend radii as small as several \SI{10}{\micro\metre} or beam splitting structures of lengths smaller than \SI{100}{\micro\metre} for total waveguide lengths of several \SI{}{\milli\metre}. 
This allows for high integration densities, which is challenging for  waveguides inscribed into glass. In contrast to systems using optical fibers and grating couplers to address the network ports, the minimum distance of our couplers is only limited by the coupler size much smaller than the diameter of an optical fiber. Thus, we can fabricate networks with a very small footprint.
As a simple example, we present Y-beam splitters and systematically investigate their fabrication parameters. 

\section{Fabrication of the Waveguides}
\begin{figure*}[hbt]
	\includegraphics{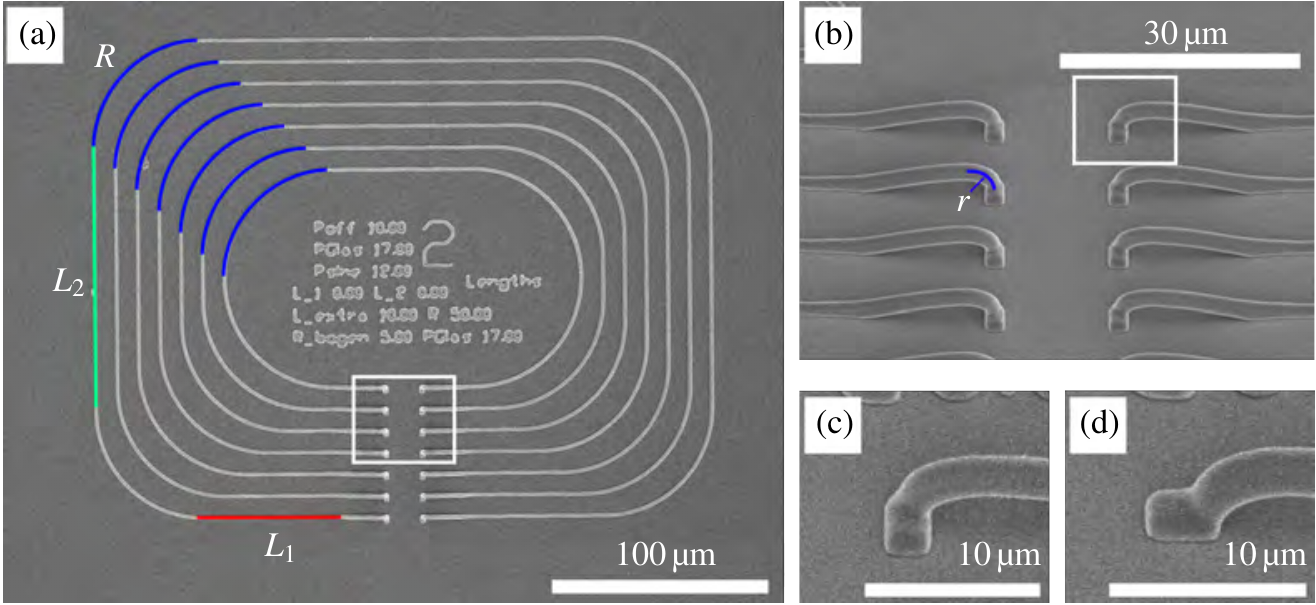}
	\caption{
		(a) SEM-image of a set of stadium waveguides with constant bend radius $R=\SI{50}{\micro\metre}$ (blue). The bend radius $R$ and lengths $L_1$ (red) and $L_2$ (green) are marked for clarity.
		(b) Zoom of white box in (a), tilted image of free standing coupling structures, the radius $r$ (blue) of the coupler is marked.
		(c) Magnification of white box in (b) showing a perpendicular coupling structure, (d) analogue for parallel coupling structure. 
	} \label{stadium-wg}
\end{figure*}
\subsection{Design Considerations}
Since our goal is to produce networks with a small footprint and moreover to integrate quantum emitters into our waveguide network, we aim for waveguides having a tight mode confinement in the waveguide.
These air-clad, bare waveguides on glass substrate are designed to have an almost rectangular cross section of approximately $\SI{1}{\micro\metre}\times\SI{1}{\micro\metre}$.
Furthermore, the waveguide layout is designed to have all ports of the network in one field of view of our microscope using perpendicular coupling structures, addressed through the glass substrate.	Because the couplers' end facets are at the interface of resist and substrate, no facet preparation is necessary.

Fig.~\ref{fib} (a)-(c) show the scanning electron microscope (SEM) image of the cross section of a typical waveguide with the simulated, guided TE-like waveguide modes. 
The intensity distribution measured at each output port depends on the specific shape of the outcoupling structure (Fig.~\ref{fib} (d)-(g)).
In the case of a coupling structure parallel to the waveguide the imaged intensity distribution resembles a Gaussian, while in the case of the coupling structure perpendicular to the waveguide the intensity distribution exhibits multiple spatial modes, probably originating from the coupler itself. 
As the transition from the waveguide to the coupling structure is not fully adiabatic for the light mode, admixtures from various spatial modes are in fact expected.

For fabrication, we use the negative tone photoresist EpoClad 50 (Microresist). The resulting waveguides comprise two types of sections.
The planar waveguide itself, guiding light parallel to the substrate surface, and the coupling structure  rotating the light's propagation direction out of plane to be perpendicular to the substrate surface, as shown in Fig.~\ref{stadium-wg}.  
The coupling structure is a quarter ring with radius $r=\SI{5}{\micro\metre}$, standing on the substrate, smoothly lowering down onto the substrate according to a $\cos^2$-function and connecting to the waveguide. 
For structure stability reasons, the radius $r$ of the incoupler is as small as \SI{5}{\micro\metre} and the transition part has a length of  \SI{20}{\micro\metre}. 
In order to increase the substrate adhesion and to improve coupling efficiency, the surface contact region is formed as coupling structures of different shapes, see Fig.~\ref{fib}(d)-(g) and Fig.~\ref{stadium-wg}(c),(d). We use shallow rectangular pads added to the end of the couplers, where we compare the performance of orientation parallel and perpendicular to the waveguide, respectively.
The waveguide section is composed of straight waveguides of lengths varying between $L_1=\SI{0}{\micro\metre}$ and $L_1=\SI{45}{\micro\metre}$, and $L_2$ between $L_2=\SI{50}{\micro\metre}$ and $L_2=\SI{150}{\micro\metre}$, connected by four quarter rings, having radius $R=\SI{70}{\micro\metre}$.
As an example, a typical set of waveguides in stadium configuration is shown in Fig.~\ref{stadium-wg}.
The waveguides feature lengths up to \SI{3.8}{\milli\metre}, bend radii $R$ (see Fig.~\ref{stadium-wg}) down to \SI{20}{\micro\metre} with a total insertion loss as low as \SI{9.1}{\decibel}, enabling high integration densities. 
The three-dimensional coupling structures have a total height of \SI{6}{\micro\metre} and a length of \SI{25}{\micro\metre}, having a total coupling efficiency of at least \SI{12}{\percent}. The insertion loss is at present dominated by in- and outcoupling loss and mode mismatch at the connection of differently shaped waveguide sections.

\subsection{Fabrication Process}

The substrates are silica coverslips (Menzel Glaeser, thickness \SI{170}{\micro\metre}), initially cleaned in an ultrasonic bath (acetone, isopropanol, deionized water, \SI{5}{\minute} each) and dried with nitrogen. 
After a short dehydration bake, the coverslips are cleaned for \SI{5}{\minute} in an oxygen plasma. 
The photoresist is dropcast onto the coverslips, baked for \SI{10}{\hour} at \SI{120}{\degreeCelsius} on a contact hotplate and cooled down natively.\\
For direct laser writing, a commercial system (Photonic Professional GT, Nanoscribe) is used with an immersion oil objective (63x, $\mathrm{NA}=1.4$). This system uses an ultrashort pulsed laser with a center wavelength of \SI{780}{\nano\metre}, a pulse length between \SI{100}{\femto\second} and \SI{200}{\femto\second}, and a repetition rate of \SI{80}{\mega\hertz}.
We typically write with speeds of \SI{50}{\micro\metre\per\second} and laser powers between \SI{6}{\milli\watt} and \SI{8.6}{\milli\watt} (in front of the objective), composing each waveguide of four parallel trajectories.
Finally, the samples are stored overnight at room temperature for post exposure bake, developed in mr-Dev 600 (Microresist), rinsed with isopropanol and dried with nitrogen.

\section{Transmission and Coupling Properties}
In order to check if the waveguides are suited for construction of networks with integrated emitters, knowledge about transmission and coupling properties is crucial. Particularly, the contributions of coupling losses, which will, for example, reduce the amount of light exciting a nano-emitter, and losses per length, which will attenuate the emitted signal of nano-emitters, are important to know. While the former can usually be compensated for by larger laser powers, the latter poses strong constraints  on, e.\,g., the level of correlations detectable.
\subsection{Experimental Setup and Methods}
\begin{figure}[hbt]
	\begin{center}
	\includegraphics{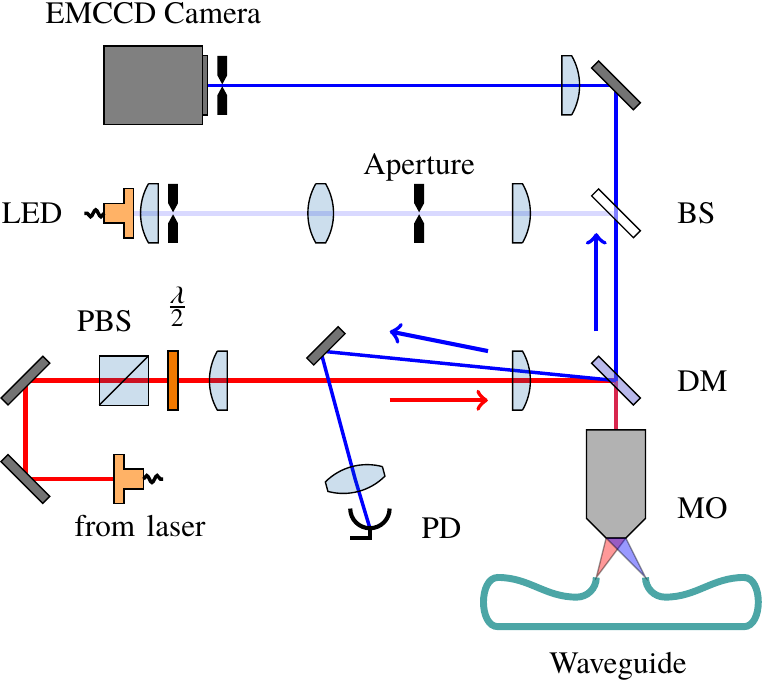}
	\end{center}
	
	\caption{Experimental setup. The microscope objective (MO) focusses the laser beam (red line) onto the input port of the waveguide under investigation. The transmitted light from the output (blue line) is collected via the same objective and sent to a photodiode (PD), the fraction transmitted at the dichroic mirror (DM) is imaged on the EMCCD camera. A polarizing beam splitting cube (PBS) and a $\lambda/2$-plate ($\lambda/2$) are used to adjust the polarization used and a beam splitter (BS) reflects the light from a LED onto the sample.}\label{aufbau}
\end{figure}

We characterize insertion loss of the waveguides and intensity distributions of the transmitted light using a homebuilt microscope (see Fig.~\ref{aufbau}). It features an objective with $\mathrm{NA}=0.65$ and a motorized stage holding the waveguide sample for scanned measurements.
\begin{figure*}[t!]
	\centering
	\includegraphics{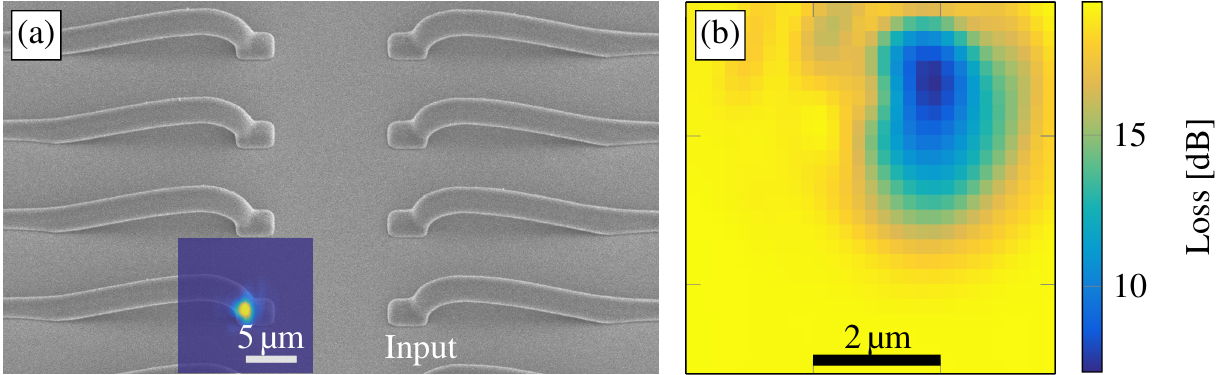}
	\caption{(a) SEM image of the coupling region of stadium waveguides, overlay: microscope image of transmitted intensity distribution, (b) incoupling position dependent transmission map of the waveguide.}
	\label{transmission}
\end{figure*}
\begin{table*}[bt]
	\begin{center}
		\setlength{\extrarowheight}{.2em}
		\begin{tabular}{|l||c|c|}
			\hline
			&$R=\SI{40}{\micro\metre}$&$R=\SI{50}{\micro\metre}$\\
			\hline\hline
			Average insertion loss & $\SI{11.93\pm0.33}{\decibel}$& $\SI{11.90\pm0.39}{\decibel}$\\
			\hline
			Maximum loss coefficient $\sigma/\Delta L$ & \SI{0.68}{\decibel \per \milli\metre}& \SI{0.81}{\decibel \per \milli\metre}\\
			\hline \hline
			Coupling loss $C$ & $\SI{11.43\pm0.98}{\decibel}$& $\SI{12.80\pm0.79}{\decibel}$\\
			\hline
			Damping $B$ & $\SI{0.96\pm 1.56}{\decibel \per \milli\metre} $& $\SI{-1.44\pm 1.22}{\decibel\per\milli\metre}$\\
			\hline
		\end{tabular}
		\caption{(Top rows) Statistical analysis of transmission data: average insertion loss and maximum loss coefficient estimated from the standard deviation $\sigma$. (Bottom rows) Fit parameters from a fit assuming linear damping on log scale for single round trip waveguides of lengths varying from \SI{0.3}{\milli\metre} to \SI{0.8}{\milli\metre}. Fit function used: $D=C+B\cdot L$ with $D$ the insertion loss, $L$ the length of the waveguide; the error is one standard deviation.
		}
		\label{no-loss}
	\end{center}
\end{table*}
In order to quantify insertion loss and transmitted mode, a single-mode infrared laserbeam ($\lambda=\SI{780}{\nano\metre}$) is focused onto the incoupling port of the waveguide. 
Each input port is scanned in a position range of approximately $\SI{4}{\micro\metre} \times \SI{4}{\micro\metre}$ by moving the sample-stage with a resolution of \SI{0.2}{\micro\metre}, while keeping the incoupling laser position fixed (see \textcolor{blue}{Visualization 1} for a video of an input-port scan).
The light transmitted is separated via a mirror and sent to a calibrated photodiode. The photodiode signals for each scan thus yield a two-dimensional image of integrated output intensities, see Fig.~\ref{transmission}. 
The insertion loss is defined as the ratio of transmitted and injected power.
In order to correct the transmission for background due to, e.\,g., scattering, the transmitted power is calculated as difference between maximum and minimum power for each scan field.

The position resolved transmitted intensity distribution for each incoupling position is obtained by imaging the waveguides' ports by an EMCCD camera (iXon3 DU-885K, Andor), see Fig.~\ref{transmission}, where also light scattered from the waveguides would be visible.

\begin{figure*}[hbt]
	\begin{center}
		\includegraphics{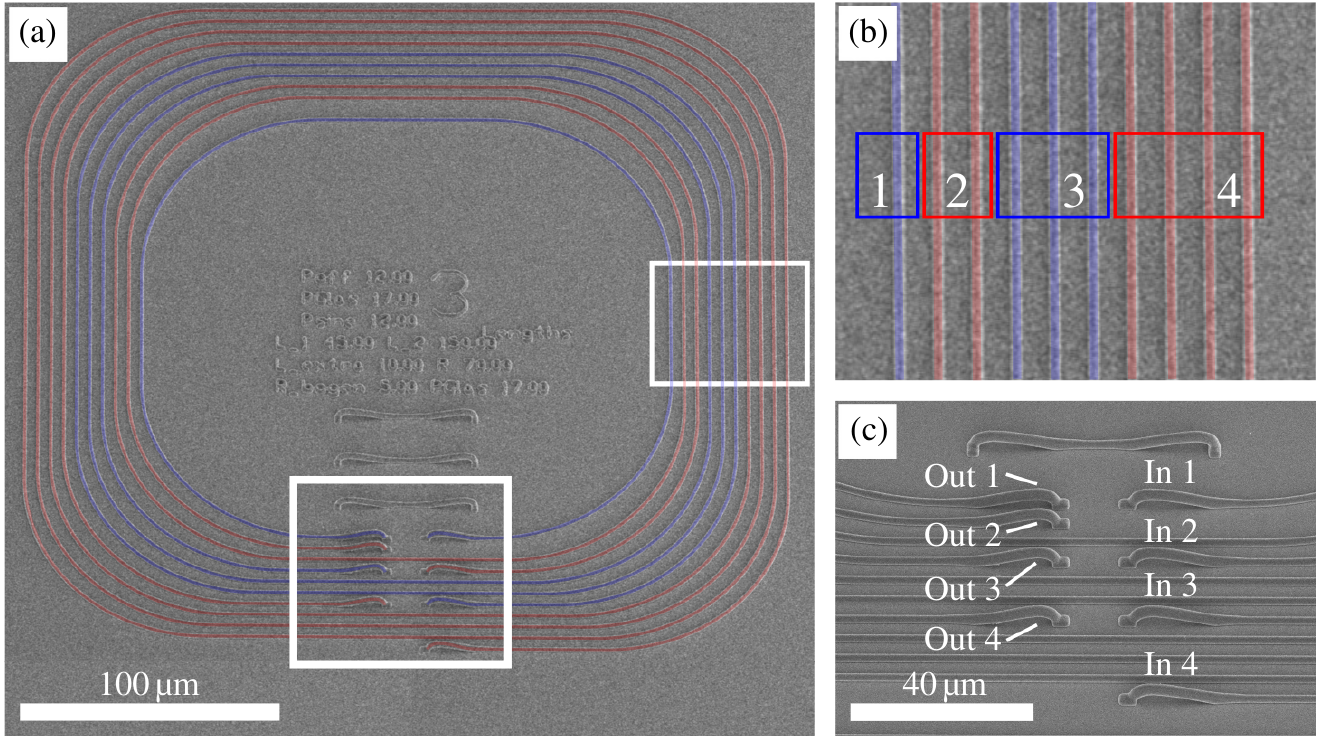}
	\end{center}
	\caption{(a) Tilted SEM image of a set of coiled stadium waveguides. Blue/red shaded: odd/even number of round trips. (b) Magnification of the sets of the parallel waveguides. (c) Magnification of the port region of the waveguides.
	}
	\label{rem_turns}
\end{figure*}
\begin{figure*}[hbt]
	\begin{center}
		\includegraphics{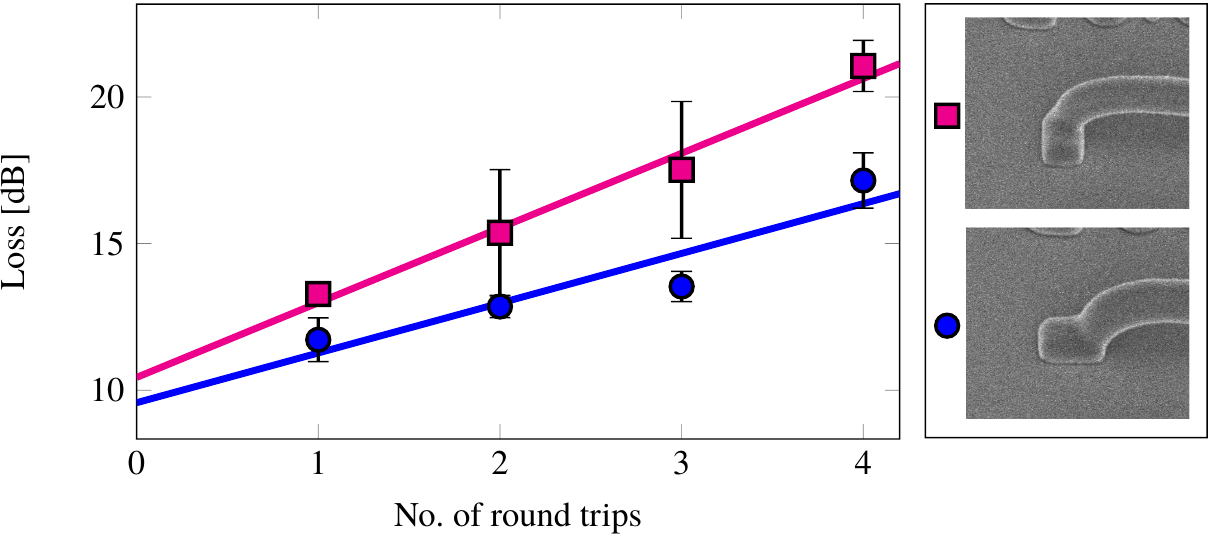}
	\end{center}
	\caption{Averaged insertion loss of coiled stadium waveguides with bars denoting standard statistical errors. The images in the legend show the corresponding SEM image of the coupling structure.
		The solid lines are linear fits with parameters given in the text.
	}
	\label{loss_length}
\end{figure*}
\subsection{Waveguide Damping and Loss Coefficients}
Insertion loss can originate from three distinct sources. 
First, losses occur at the in- and outcoupling ports due to mode mismatch with the free space beam and due to the small coupler geometry.
Second, absorption or scattering in the waveguides can lead to a constant loss per unit length, while, third, at each interface between straight and bent waveguide parts losses can occur due to improper mode matching.
In order to estimate the contributions of loss per length, loss due to the bend radius, and loss due to imperfect mode-matching, we compare three different sets of waveguides. 
First we extract the length dependent loss coefficient and the loss due to coupling through variation of the waveguide length similar to the cutback technique.

\begin{figure*}[hbt]
\centering
	\includegraphics{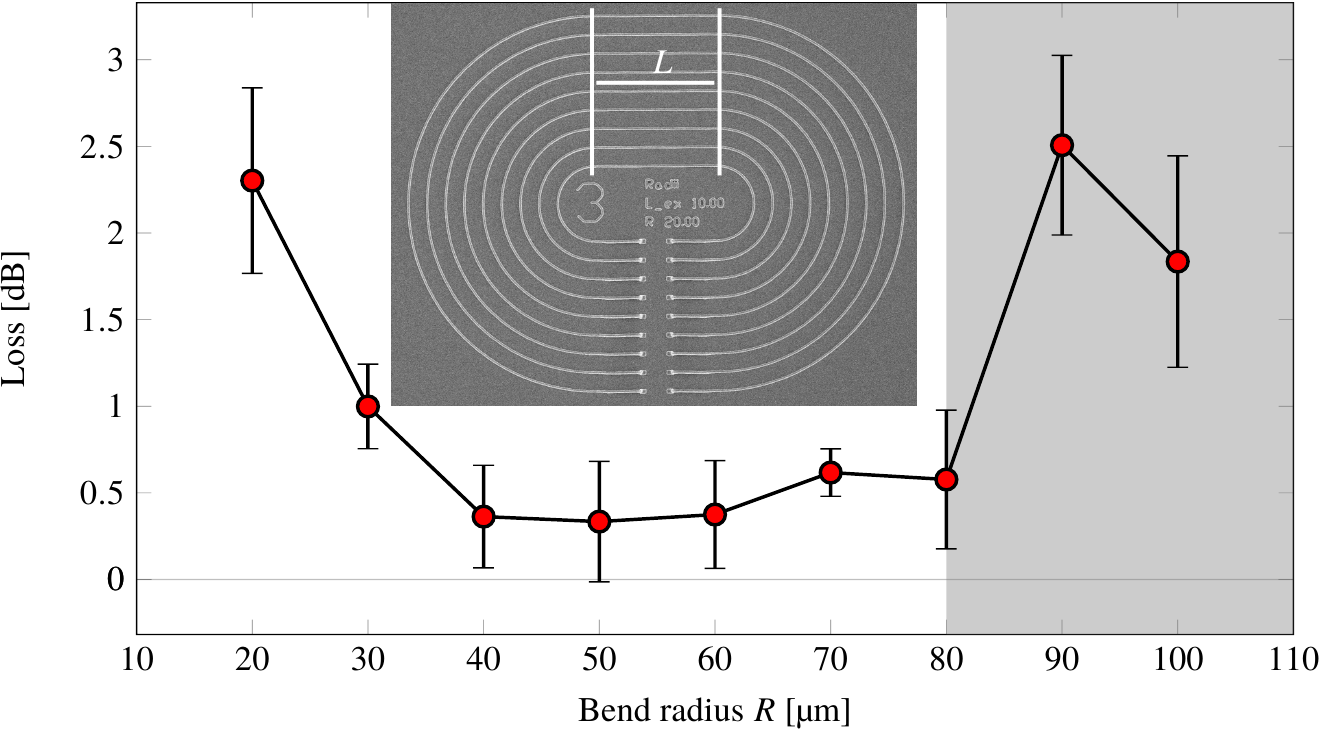}
	\caption{Average bend dependent loss, the error bars denote standard statistical errors. The inset shows a SEM image of one field containing the structures under investigation with constant length $L=\SI{70}{\micro\metre}$ and varying bend radius $R$.
	The gray area marks the unexpected increase in waveguide loss when increasing the bend radius.
	}
	\label{loss_radii}
\end{figure*}

Therefore, in a first set of waveguides, shown in Fig.~\ref{stadium-wg}, the lengths $L_1$ and $L_2$ of the waveguides are increased for two different curvature radii $R$, while keeping the number of straight-to-bent transitions constant. 
Within our measurement uncertainty we do not observe an increased loss, neither when increasing the total waveguide length by a factor of three to \SI{0.9}{\milli\metre}, see Tab.~\ref{no-loss}, nor when changing the bend radius from $R=\SI{40}{\micro\metre}$ to $R=\SI{50}{\micro\metre}$, consistent with Fig.~\ref{loss_radii}. 
The transmission data taken does not show any obvious trend. 
We thus give the average insertion loss and estimate an upper bound for the loss per unit length by using our measurement uncertainty as maximum loss for the longest waveguide. Thus the loss per unit length is smaller than \SI{0.68}{\decibel\per\milli\metre} for a bend radius of $R=\SI{40}{\micro\metre}$ and smaller than $\SI{0.81}{\decibel\per\milli\metre}$ for $R=\SI{50}{\micro\metre}$. 
Alternatively, we fit linear damping on a logarithmic scale (see Tab.~\ref{no-loss} for the parameters), which yields reasonable values for the coupling loss, but the parameters for damping are compatible with zero damping, or even show negative damping.

Since the damping coefficient is not reliably determined from this set of waveguides, the length of the waveguides is increased even more by coiling up the waveguides with fixed curvature radius $R$, keeping the dimensions below $\SI{300}{\micro\metre}\times\SI{300}{\micro\metre}$, the range accessible by the Nanoscribe's piezo stage. 
One field containing a set of these test structures is shown in Fig.~\ref{rem_turns}.
We fabricate five identical sets of these fields. Since two of the fields do not guide light properly due to fabrication imperfections, such as improper automated localization of the interface between coverslip and photoresist during DLW, we evaluate the other three fields in the following.
The total lengths of the waveguides shown are \SI{0.61}{\milli\metre}, \SI{1.42}{\milli\metre}, \SI{2.46}{\milli\metre}, and \SI{3.85}{\milli\metre} respectively.
The losses observed are shown in Fig.~\ref{loss_length} and are proportional to the number of turns or, equivalently, number of transitions from straight to bent waveguide parts and vice versa. To illustrate the effect of different coupler geometries, we compare the results for two differently oriented rectangular couplers.
For parallel coupling structures we get an extrapolated combined in- and outcoupling loss of \SI{9.57}{\decibel} and a loss per round trip of \SI{1.67}{\decibel}, for perpendicular coupling structures the in- and outcoupling loss is \SI{10.43}{\decibel} and the loss per round trip is \SI{2.55}{\decibel}.
The incoupling loss for both structures coincides but the loss per round trip differs, which is likely due to the perpendicular coupling structure itself causing increased loss for the adjacent waveguide.\\

The fabrication of waveguide networks with high integration density requires small bend radii, limited eventually by losses of the light mode from the bent waveguide.
To investigate such bend induced loss, we study a third set of structures and vary the bend radius $R$ from $R=\SI{20}{\micro\metre}$ to $R=\SI{100}{\micro\metre}$.
One field containing these waveguides is shown in Fig.~\ref{loss_radii}. 
Again, five copies of this field are fabricated and all of them are evaluated.
The data from different writing fields containing the structures shown in Fig.~\ref{loss_radii} show the same scaling with bend radius $R$, however, they differ by an offset against each other. Therefore, we subtract the minimum insertion loss measured in one field from all waveguides in this respective field, to extract the bend induced loss.
For radii between $R=\SI{20}{\micro\metre}$ and $R=\SI{80}{\micro\metre}$, the bend loss shows the expected behavior: it decreases with increasing radius up to $R=\SI{40}{\micro\metre}$ by approximately \SI{2}{\decibel} and shows approximately no further change for larger radii.
Above $R=\SI{80}{\micro\metre}$, the bend loss increases again. Since this behavior is not expected from waveguide theory, this part of Fig.~\ref{loss_radii} is shaded in gray.
Possible reasons for this unexpected increase can include tensions in the structure or too small substrate adhesion, leading to local peeling.
Since we are interested in small bend radii, this is not limiting for us and not further investigated.
For all other measurements shown in this work, a bend radius of $R=\SI{70}{\micro\metre}$ is used.

We conclude from these measurements that the dominant source of losses originates from imperfect mode matching between waveguide sections, where the measured value depends on the specific design of waveguide and coupling structure. Also a small foot-print of optical elements is possible with bend radii down to \SI{40}{\micro\metre}. 
These results are promising for constructing quantum optical networks with high integration density, where the in-coupling losses can be easily compensated for, while sensitive signals can be guided with low loss.

\section{Fluorescence Measurements}
\begin{figure}[h!bt]
	\begin{center}
	\includegraphics{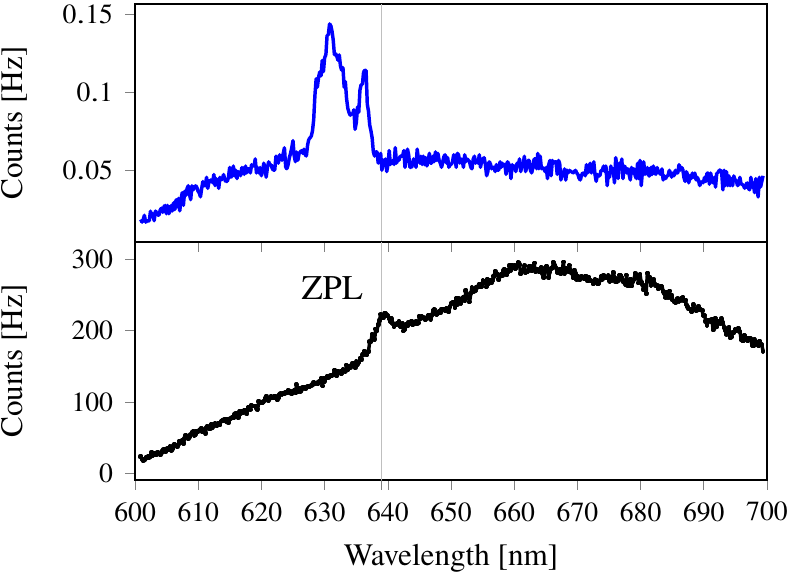}
	\end{center}
	\caption{Comparison of fluorescence spectra of EpoClad 50 (blue curve, excitation power \SI{117}{\micro\watt}) and a nano-diamond containing approximately 1200 NVC (black curve, \SI{470}{\micro\watt}), excited at $\lambda=\SI{532}{\nano\metre}$.
	}
	\label{fluorescence}
\end{figure}
\begin{figure*}[hbt]
	\includegraphics{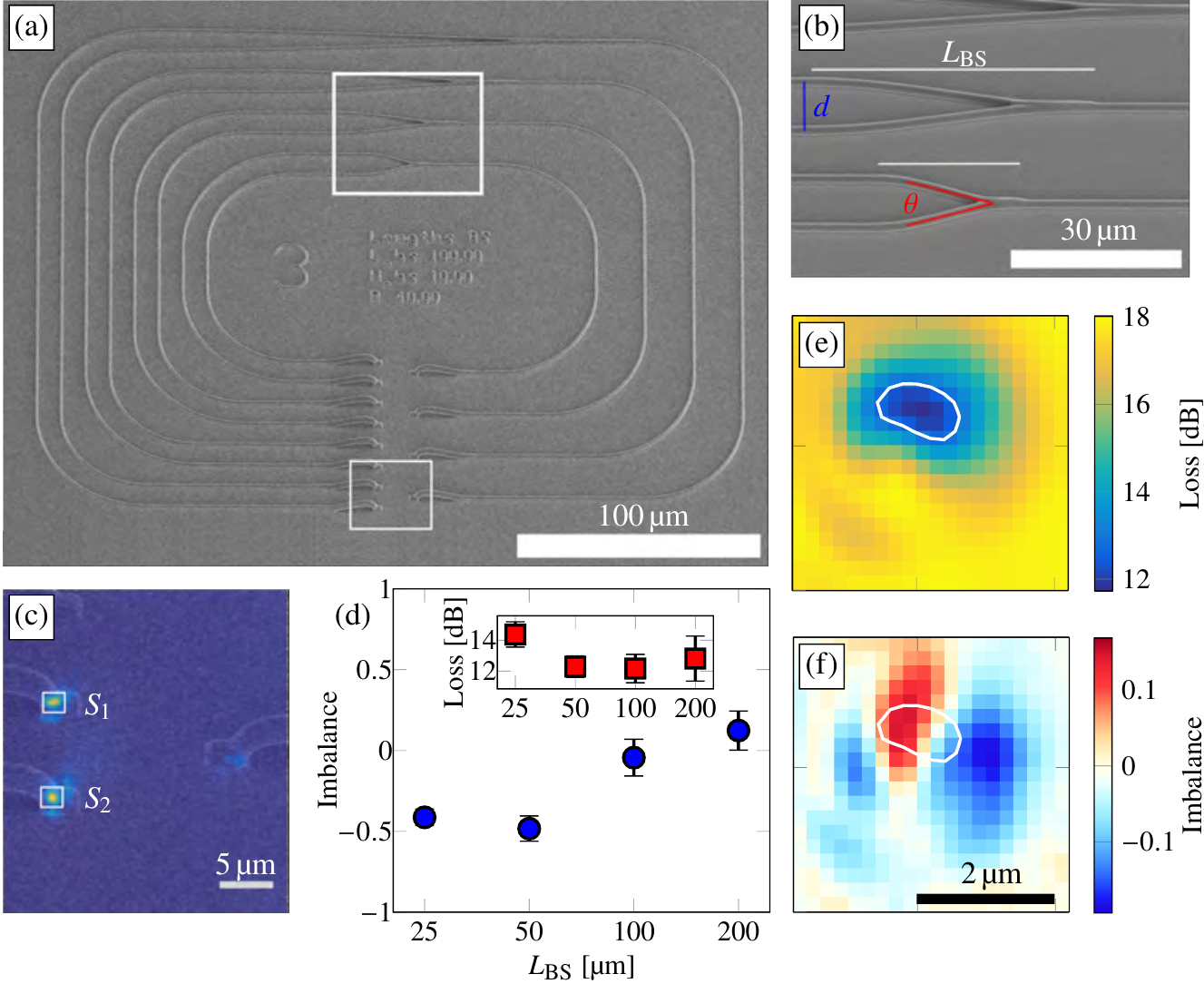}
	\caption{(a) Tilted SEM image of a set of Y-splitters with varying length $L_\mathrm{BS}$,
		(b) magnification of the splitting region. The splitting angle $\theta$ of the waveguides is varied by changing the splitter length $L_\mathrm{BS}$, 
		(c) magnification of the input and output region overlaid with the detected intensity distribution at maximum transmission, (d) averaged imbalance and insertion loss of the splitters as a function of $L_\mathrm{BS}$, or, equivalently, the splitting angle $\theta$. Two-dimensional maps of (e) insertion loss and (f) imbalance of the longest waveguide shown as a function of the incoupling position. The device transmits at least \SI{80}{\percent} of its maximum transmission in the area enclosed by the white line. 
	}
	\label{rem_y_bs}
\end{figure*}
For the future we aim on integrating single quantum emitters, such as NVC in nano-diamonds, into our waveguide network. 
The signal of these single emitters might be masked due to fluorescence from waveguides or the substrate itself.
Therefore we investigate the fluorescence of the photoresist used, EpoClad 50, upon excitation at \SI{532}{\nano\metre}.
The resulting spectrum is shown in Fig.~\ref{fluorescence} in comparison to the emission of a single nano-diamond containing many NVC ($>$1200 NVC/particle, fluorescent nano-diamond, Sigma-Aldrich).
The resist spectrum shows two distinct peaks at wavelengths slightly below the zero-phonon-line of the NVC at \SI{638}{\nano\metre} on top of a flat and broad background. Integrating the total fluorescence signal, we find a resist fluorescence which is only 60\% above the fluorescence level of the glass substrate used.
In order to directly compare the signals from resist with NVC fluorescence, we also integrate the NVC spectrum\cite{footnote}, normalize the signals with respect to the respective excitation power and the NVC signal to the estimated number of 1200 NVC in the nano-diamond investigated. We find total fluorescence levels in the same order of magnitude for resist and a single NVC.
In future work, single NVC will be integrated into the waveguide network. In fact,
single emitters integrated into waveguides with a refractive index near to the substrate as well as their characterization were reported in \cite{Shi.2016}.
There, it has been shown that the integration into waveguides efficiently guiding the emitted single photons increases the signal to noise ratio, facilitating coupling to single emitters also in our waveguides.

\section{Towards Complex Networks}
Beam splitters are the most important optical element to construct waveguide networks. We fabricate and evaluate five copies of the set of Y-beam splitters shown in Fig.~\ref{rem_y_bs},  as a simple realization for our waveguide system. The splitters differ from each other by the splitting angle, realized by varying the length over which the splitted waveguides separate to the final distance $d=\SI{10}{\micro\metre}$, see Fig.~\ref{rem_y_bs}(b). The splitters shown have angles $\theta=\SI{33.9}{\degree}, \SI{18.1}{\degree}, \SI{8.8}{\degree}, \SI{4.5}{\degree}$. In order to characterize the performance, the transmitted power of the whole splitter is again recorded by a photodiode as a function of input port position,  yielding a two-dimensional transmission map for each output port. From these maps, a two-dimensional loss map is calculated, see Fig.~\ref{rem_y_bs}(e). 
In order to estimate the imbalance for each incoupling position, we take the camera image Fig.~\ref{rem_y_bs}(c) and calculate the imbalance $I$ as
\begin{equation}\label{eq:imbalance}
I=\frac{C_\mathrm{1}-C_\mathrm{2}}{C_\mathrm{1}+C_\mathrm{2}},
\end{equation}
where $C_1$, $C_2$ are the integrated intensities in the white squares $S_1$, $S_2$ in Fig.~\ref{rem_y_bs}(c). This yields a two-dimensional map of the splitter's imbalance, see Fig.~\ref{rem_y_bs}(f).
We identify the range of incoupling positions where the transmission is above \SI{80}{\percent} of the maximum transmission and calculate for these pixels the average imbalance as well as its standard deviation shown in Fig.~\ref{rem_y_bs}(d).

The insertion loss decreases with increasing beam splitter length and also the beam splitters become more balanced.
A reason for the splitters' imbalance might originate from a small structural asymmetry resulting from the writing process. This limitation, however, can be resolved through parameter optimization.

These results show that waveguide networks can be fabricated from the presented photoresist material system. By concatenating the  beam splitter design shown with adjusted splitting angles, the power distribution in the network's branches might be adjusted.

\section{Conclusions}
In summary, we have demonstrated up to \SI{3.8}{\milli\metre} long direct laser written waveguides with an insertion loss as low as \SI{9.1}{\decibel}, showing low loss for bend radii between  $R=\SI{40}{\micro\metre}$ and $R=\SI{80}{\micro\metre}$.
The long waveguide itself is laid down onto the silica substrate, while for in- and outcoupling of light, three dimensional couplers are used. 
These couplers are viewed simultaneously with a single microscope objective, giving access to the whole network while leaving the other side of the substrate open for further manipulation. 
As a first device, we show simple Y-beam splitters with an insertion loss of \SI{12}{\decibel} and an adjustable splitting ratio via variation of the splitting region's length.

Our next step is to integrate nano-diamonds containing NVC into our structures, which is expected to increase the signal-to-noise ratio \cite{Shi.2016}.
For our photoresist, we find an autofluorescence of the same order as the NVC fluorescence.
Further, we aim at scaling up the size and complexity of our network and showing interferometers and resonators. This will give access to, e.\,g., the phase and frequency of the transmitted photons.
The versatility of the system used also facilitates combining our waveguide networks with existing photonic chips or connecting the waveguides directly to detectors, e.\,g. photodiodes, or light sources.
The polarization of light transmitted through the waveguide network is an important degree of freedom; DLW allows access to control all waveguide parameters to tune, for instance, the polarization of the propagating light.
\section{Acknowledgments}
A. Landowski is a recipient of a DFG-fellowship through
the Excellence Initiative by the Graduate School Materials Science in Mainz (GSC 266).
We thank Michael Renner for initial support with the DLW process and Michael Schmidt for technical support in the initial phase of the project. 
We also acknowledge technical support by the Nano Structuring Center, TU Kaiserslautern.


\bibliographystyle{osajnl}

\begin{thebibliography}{10}
\bibitem{footnote}
	The NVC spectrum does not change for a single NVC, since the phonon broadening is homogeneous.
\end{thebibliography}
\end{document}